\date{}
\documentclass[11pt]{article}
\usepackage{hyperref}
\usepackage{tikz}

\usepackage{amsmath,amsthm,amsfonts,amssymb,amsopn,amscd} 
\usepackage{color}

\def\qed{{\unskip\nobreak\hfil\penalty50
\hskip2em\hbox{}\nobreak\hfil$\square$
\parfillskip=0pt \finalhyphendemerits=0\par}\medskip}
\def\proof{\trivlist \item[\hskip \labelsep{\bf Proof.\ }]}
\def\endproof{\null\hfill\qed\endtrivlist\noindent}

\def\act{\underset{s\to -it}{\textnormal {anal.cont.\,}}}

\def\tilde{\widetilde}

\def\id{{\rm id}}

\def\a{\alpha}
\def\b{\beta}

\def\e{\varepsilon}

\def\Ga{\Gamma}

\def\l{\lambda}

\def\r{\rho}

\def\om{\omega}
\def\Om{\Omega}

\def\proof{\trivlist \item[\hskip \labelsep{\bf Proof.\ }]}
\def\endproof{\null\hfill\qed\endtrivlist\noindent}
\def\eproof{\null\hfill\qed\endtrivlist\noindent}

\newcommand{\bthm}{\begin{theorem}}
\newcommand{\ethm}{\end{theorem}}

\newcommand{\bprop}{\begin{proposition}}
\newcommand{\eprop}{\end{proposition}}
\newcommand{\bcor}{\begin{corollary}}
\newcommand{\ecor}{\end{corollary}}
\newcommand{\blem}{\begin{lemma}}
\newcommand{\elem}{\end{lemma}}

\def\setminus{\smallsetminus}

\def\A{{\cal A}}

\def\F{{\cal F}}

\def\M{{\cal M}}

\def\R{{\cal R}}

\def\H{{\cal H}}
\def\K{{\cal K}}
\def\S{{\cal S}}

\def\f{{\varphi}}
\def\s{{\sigma}}

\def\l{{\lambda}}

\def\PSL{{{\rm PSL}(2,\mathbb R)}}

\def\S2{S^{1(2)}}

\def\ov{\overline}
\def\RR{\mathbb R}

\def\sr{\textit{\small R}}
\newtheorem{theorem}{Theorem}[section]
\newtheorem{lemma}[theorem]{Lemma}

\newtheorem{corollary}[theorem]{Corollary}

\newtheorem{proposition}[theorem]{Proposition}

\theoremstyle{definition} 

\theoremstyle{remark} \newtheorem{remark}[theorem]{Remark}

\newcommand{\ben}{\begin{equation}}
\newcommand{\een}{\end{equation}}

\def\setminus{\smallsetminus}

\def\PSL{PSU(1,1)}

\def\CC{{\mathbb C}}

\def\SL2{{{\rm SL}(2,\R)}}

\def\PSL2{{{\rm PSL}(2,\Reali)}}

\def\U1{{{\rm V}(1)}}
\def\SU2{{{\rm SV}(2)}}

\def\SU{{{\rm SU}}}

\def\A{{\mathcal A}}

\def\F{{\mathcal F}}
\def\H{{\mathcal H}}
\def\L{{\mathcal L}}

\def\K{{\mathcal K}}

\def\M{{\mathcal M}}

\def\P{{\mathcal P}}

\title{\Huge{Bekenstein
Bound for Approximately Local Charged States
}}

\author{ {\sc Stefan Hollands}\\
Institut für Theoretische Physik, Universität Leipzig \\
Brüderstrasse 16, 04103 Leipzig, Germany\\
Max Planck Institute for Mathematics in Sciences (MiS)\\
Inselstra{\ss}e 22, 04103 Leipzig, Germany
\\
{}
\\
{\sc Roberto Longo}\\
Dipartimento di Matematica,
Tor Vergata Universit\`a di Roma\\
Via della Ricerca Scientifica, 1, I-00133 Roma, Italy
}

\date{}
\begin{document}

\maketitle

\begin{abstract}
We generalize the energy-entropy ratio inequality in quantum field theory (QFT) established by one of us
from localized states 
to a larger class of states. The states considered in this paper can be in a charged (non-vacuum) representation of the QFT
or may be only approximately localized in the region under consideration. Our inequality is $S(\Psi |\!| \Omega) \le 2\pi R \,
    ( \Psi, H_\rho \Psi ) + \log d(\rho) + \varepsilon$, where $S$ is the relative entropy, where $R$ is a ``radius'' {(width)} characterizing the size of the region, $d(\rho)$ is the statistical (quantum) dimension
    of the given charged sector $\rho$ hosting the quantum state $\Psi$, $\Omega$ is the vacuum { state}, $H_\rho$ is the Hamiltonian in the charged sector, and $\varepsilon$
    is a tolerance measuring the deviation of $\Psi$ from the vacuum according to observers in the causal complement of the region.
\end{abstract}

\newpage

\section{Introduction}
In a recent paper \cite{L24}, one of us proved the inequality
\begin{equation}
\label{bekenstein1}
    S(\Phi |\!| \Omega)_{\A(B)} \le 2\pi R \,
    ( \Phi, P^0 \Phi )
\end{equation}
between the relative entropy associated with the observable algebra $\A(B)$ of a region, $B$, of a time-slice in Minkowski spacetime, and the expectation value of the Hamiltonian, $P^0=H$, generating time-translations in the direction orthogonal to the time-slice. In \cite{L24}, the vector state $\Phi$ in \eqref{bekenstein1} was assumed to be of the form $\Phi = v\Omega$, where $v$ is an isometry in the local algebra of observables associated with $B$, and $\Omega$ is the vacuum state of the quantum field theory (QFT). $R$ is a measure of the size of the region $B$, defined precisely as the minimum distance between two parallel planes enclosing $B$ in the time-slice. 

Although the quantity on the left of \eqref{bekenstein1} is the relative entropy, a measure of the distinguishability of $\Phi, \Omega$ from the point of view of $B$ -- and not the entropy itself, it is reminiscent of an inequality proposed by Bekenstein \cite{B81}. Therefore, despite this important conceptual difference and as in \cite{L24}, we shall refer to \eqref{bekenstein1} as the ``Bekenstein bound''.\footnote{For an earlier but different interpretation of the Bekenstein bound in QFT involving the boost Hamiltonian, rather than the generator $P^0$ of time-translations as in \eqref{bekenstein1}, see \cite{Ca08}. For an interpretation of \cite{Ca08} involving a crossed-product construction, see
\cite{Kudler-Flam:2023hkl}.} 

The states $\Phi = v\Omega$ appearing in \eqref{bekenstein1} are called ``localized'', because {they are characterized by the property}  $(\Phi, b'\Phi) = (\Omega, b'\Omega)$ for any $b' \in \A(B)'$, the commutant algebra of the algebra of observables $\A(B)$ associated with $B$. Localized states form an important subclass including, e.g., coherent states in the QFT of a free boson, but are quite special from the viewpoint of entanglement theory. It is therefore natural to look for generalizations of \eqref{bekenstein1} to broader classes of states $\Phi$.
In this work, we will focus on the following: 

\begin{enumerate}
    \item A class of vector states $\Phi$ in the vacuum representation such that $\Phi$ is not necessarily of the form $v\Omega$ for some isometry $v$ associated with the region $B$, but which are, in a sense, close to the vacuum in the causal complement of $B$. 

    \item A class of vector states outside the vacuum representation of the QFT. The states that we consider are related to charged sectors localized in $B$, and are equal to the vacuum in the causal complement of $B$.
\end{enumerate}

We will also consider a combination of 1) and 2), i.e., states carrying charges within $B$ that are not precisely equal but close to the vacuum state in the causal complement of $B$, see Theorem \ref{Thme2}. 

When considering the Bekenstein bound \eqref{bekenstein1} for more general states $\Phi$ than those of the form $\Phi=v\Omega$ considered in \cite{L24}, it is important to bear in mind that the relative entropy in \eqref{bekenstein1} only depends on the restriction of the expectation value functional $b \mapsto (\Phi, b\Phi)$ to observable {algebra} $\A(B)$. Thus, the relative entropy remains unchanged passing from $\Phi$
to the vector state $\Phi' = v'\Phi$, for any isometry $v' \in \A(B)'$. Since, by choosing a $v'$ with an enormous energy transfer, we may make the right side $(\Phi', P^0\Phi')$ of \eqref{bekenstein1} arbitrarily large, the Bekenstein bound does not have much content unless we impose some conditions on the vector representative $\Phi'$ as a state functional $b' \mapsto (\Phi', b'\Phi')$ on observables $b' \in \A(B)'$.  

Our central idea for 1) is therefore to consider the interplay between the complementary relative entropies, $S(\Phi |\!| \Omega)_{\A(B)}$ and $S(\Phi |\!| \Omega)_{\A(B)'}$. To this end, we will first derive a sum rule\footnote{Proposition 3.2 can be seen as a generalization of \cite[Lemma 1]{C18}.} (Proposition \ref{SSP}) relating these with $(\Phi, P^0 \Phi)$. Then, we will consider a restriction on $\Phi$ making $S(\Phi |\!| \Omega)_{\A(B)'}$
small. Our restriction is to demand that the expectation value functional $\A(B)' \owns b' \mapsto (\Phi, b' \Phi)$
is dominated by $(1+\varepsilon)$ times 
$\A(B)' \owns b' \mapsto (\Omega, b' \Omega)$, where we think of $\varepsilon$ as small. Such a condition is equivalent to demanding that the order $p=\infty$
sandwiched Renyi divergence $D(\Phi |\!| \Omega)_{\A(B)'}$ is less than $\log (1+\varepsilon)$, see e.g., \cite{Tomamichel}. As is known, the order $p=\infty$  Renyi divergence may be viewed as a ``1-shot'' measure of distinguishability between $\Phi, \Omega$, which dominates the Araki-Umegaki relative entropy $S(\Phi |\!| \Omega)_{\A(B)'}$, an ``asymptotic''\footnote{I.e., one requires an infinte number of copies of the system identically prepared in the state $\Phi$ with respect to $B$.} measure from an operational viewpoint, see e.g., \cite{Wilde}. Thereby, we find that our domination condition implies the Bekenstein bound \eqref{bekenstein1}
up to a tolerance $\varepsilon$, see Theorem \ref{Thme}. We think it is physically reasonable to impose a proximity of $\Phi$ to $\Omega$ from the viewpoint of the environment, $B'$, in terms of a 1-shot divergence, since one cannot prepare arbitrarily many copies of a state associated with an infinite environment. 

Furthermore, if a {conjecture} due to Connes-Takesaki \cite{CT77} about the state space of type $III_1$ von Neumann algebras is true, then vectors $\Phi$ which induce a given expectation value functional on $\A(B)$ and at the same time satisfy the domination condition on $\A(B)'$ can be found for as small an $\varepsilon$ as we like, see Remark \ref{rem3.6}. 

For 2), we consider localized vector states in a charged representation of the QFT. In the algebraic approach to QFT used in this paper, such states basically correspond to expectation 
functionals $\A(B) \owns b \mapsto (\Phi, \rho(b)\Phi)=:\varphi(b)$, where $\rho$ is a DHR endomorphism  of $\A(B)$, and where $\Phi = v\Omega$ is an ordinary localized state as appearing in \eqref{bekenstein1}. The endomorphism $\rho$ plays the role of  representation in the DHR formalism, see e.g., \cite{H}. A central concept in this formalism is the notion of a ``left-inverse'', $\Psi_\rho: \A(B) \to \A(B)$, which is a completely positive, unital, linear map characterized by: $\Psi_\rho \rho = {\id}$, and $E_\rho = \rho \Psi_\rho$ is a so-called conditional expectation from $\A(B) \to \rho[\A(B)]$. It is well-known that the relative entropy plays well with conditional expectation values and completely positive maps, more generally. In fact, using the concept of left inverse, we shall show (see Theorem \ref{Bch2} for the precise statement and assumptions) that
\begin{equation}
\label{bekenstein3}
    S(\varphi |\!| \omega)_{\A(B)} \le 2\pi R \,
    ( \Phi, P^0_\rho \Phi ) + \log d(\rho),
\end{equation}
where $\omega(b) := (\Omega, b\Omega)$ is the expectation functional on $\A(B)$ induced by the vacuum vector $\Omega$.
Compared with \eqref{bekenstein1}, 
we have the Hamiltonian $P^0_\rho$ in the representation $\rho$ [heuristically, $P^0_\rho = \int \rho[T^0{}_0(0,x)] d^3x$, where $T_{\mu\nu}$ is the stress energy tensor of the QFT], and we have the logarithm of the DHR statistical dimension, $d(\rho)$, defined as the smallest positive real number such that {$E_\rho(b^*b) \ge d(\rho)^{-2} b^* b$} for all $b \in \A(B)$ {(Pimsner-Popa bound for the Jones index, see \cite{L89})}.
The statistical dimension is indicative of the size of the multiplet of charged localized fields generating the charged state from the vacuum. It is natural for its logarithm to appear on the right side of \eqref{bekenstein3} because the size of the multiplet effectively corresponds to a kind of multiplicity (the dimension of the representation under which the charged fields transform). 

This paper is organized as follows. In Section \ref{sec1} we provide the general technical background 
and some useful auxiliary results for the remainder of the paper. 
Section \ref{Appr} contains our main results, and in Section \ref{outlook} we give an outlook on some interesting, in our opinion, follow-up questions. 

\section{General background}
\label{sec1}
In this section we set up {and recall} the basic mathematical  structure needed for our results. 

\subsection{Unitary families and Leibniz rule}
If $z(s)$ is a complex-valued function on a real interval $(-a, a)$, and $\a,\b\in \RR\cup \{\pm \infty\}$,we write 
\ben\label{gl}
\lim_{s\to 0} z(s) = \a + i\b
\een
if $\lim_{s\to 0} \Re z(s) = \a$ and $\lim_{s\to 0} \Im z(s) = \b$. Then we put
\ben\label{gl1}
\partial_s z(s)   = \lim_{s\to 0}\frac{z(s) - z(0)}s
\een
if the limit of the incremental ratio exists in the above sense. 

Let $V(s)$, $s\in (-a,a)$, $a >0$, be a familiy of unitaries on the Hilbert space $\H$, and $\Phi\in\H$ a unit vector. Note that
\ben\label{iff}
||V(s)\Phi - \Phi||^2 = o(s)\quad {\rm iff} \quad  \partial_s\Re(\Phi, V(s)\Phi) = 0\, .
\een
This follows because $||V(s)\Phi - \Phi||^2 = 2 - 2\Re\big(\Phi, V(s)\Phi)$, so
\[
||V(s)\Phi - \Phi||^2 = -2\Re\frac{(\Phi, V(s)\Phi) - 1}{s}s\, .
\]
Therefore, we have the following lemma.
\blem\label{Ds}
Let $V(s)$, $s\in (-a,a)$, $a >0$, be a family of unitaries on the Hilbert space $\H$, and $\Phi\in\H$ a unit vector. Then $-i\partial_s (\Phi, V(s)\Phi)  = \a \in\RR\cup \{\pm\infty\}$ iff
\[
\lim_{s\to 0}\frac{\Im(\Phi, V(s)\Phi)}{s}  =  \a\, ,\quad ||V(s)\Phi - \Phi||^2 = o(s)\, .
\]
\elem
\proof
Immediate by the above comments. 
\eproof
Let $A$ be a selfadjoint operator on the Hilbert space $\H$ and $\Phi\in\H$ a vector. We say that $(\Phi, A\Phi)$ is {\it well-defined} if either 
$||A^{1/2}_+\Phi||^2 < \infty$ or $||A^{1/2}_-\Phi||^2 < \infty$, where $A_\pm$ is the positive/negative part of $A$,
and in this case we set 
\[
(\Phi, A\Phi) = ||A^{1/2}_+\Phi||^2  -  ||A^{1/2}_-\Phi||^2;
\]
we say that $(\Phi, A\Phi)$ is {\it finite} if $ ||\, |A|^{1/2}\Phi || < \infty$, that is, $||A^{1/2}_+\Phi||^2 < \infty$ and $||A^{1/2}_-\Phi||^2 < \infty$. 
\blem\label{Ad}
Let $A$ be a selfadjoint operator on the Hilbert space $\H$, $U(s) =  e^{isA}$, and $\Phi\in \H$ a vector. If  $(\Phi, A\Phi)$ finite,
we have
\[
-i \partial_s (\Phi, U(s)\Phi)  
= (\Phi, A\Phi) \, ,
\]
that is,
\ben\label{infA}
\lim_{s\to 0} \frac{\Im (\Phi, U(s)\Phi)}{s} 
= (\Phi, A\Phi)\, , \qquad  ||U(s)\Phi - \Phi||^2 = o(s) \, .
\een
\elem
\proof
We may assume $||\Phi|| = 1$. 
Let $\mu$ be the spectral probability measure on $\RR$ of $A$ associated with $\Phi$.  
We have
\[
\frac{\Im (\Phi, U(s)\Phi)}{s}  =  \int_{-\infty}^\infty \frac{\Im e^{i\l s} }{s}d\mu(\l)
=   \int_{-\infty}^\infty \frac{\sin(\l s)}{ s}  d\mu(\l)  \, .
\]
Now, $\left|\frac{\sin(\l s)}{s} \right| \leq |\l|$, and $\int_{-\infty}^\infty |\l|d\mu(\l) < \infty$ because $(\Phi, A\Phi)$ is finite; so,
by the dominated convergence theorem, we have
\[
\lim_{s\to 0}  \int_{-\infty}^\infty \frac{\sin(\l s)}{s}  d\mu(\l)  =  \int_{-\infty}^\infty \l d\mu(\l) =  (\Phi, A  \Phi) \, ,
\]
that shows the first equality in \eqref{infA}. 

Concerning the second limit in \eqref{infA}, note that
\[
\Re\frac{(\Phi, U(s)\Phi) -1}{s}  =  \int_{-\infty}^\infty \Re\frac{e^{i\l s} - 1}{s}d\mu(\l)
=  \int_{-\infty}^\infty \frac{\cos(\l s) - 1}{ s}  d\mu(\l) \, .
\]
Then, similarly as above,
\[
 \lim_{s\to 0}\int_{-\infty}^\infty \frac{\cos(\l s) - 1}{ s}  d\mu(\l) = 0 \, ,
\]
 so $\lim_{s\to0}\Re\big((\Phi, U(s)\Phi) -1\big)/s = 0$.
 Finally, note that
 $
 ||U(s)\Phi - \Phi||^2  = o(s) $
by  \eqref{iff}.  
\eproof
\bprop\label{Up1}
Let $w_k(s)$, $s\in (-a,a)$, $a> 0$, be families of unitaries on the Hilbert space $\H$, and $\Phi\in\H$ a vector state such such that $-i\partial_s (\Phi, w_k(s)\Phi)  = \a_k \in\RR$, $k=1,2, \dots, n$. 

Then
\ben\label{UV}
-i\partial_s (\Phi, Z(s)\Phi)  = \a_1 + \a_2 +  \cdots +\a_n\, ,
\een
with $Z(s) = V_1(s)V_2(s)\dots V_n(s)$. 
\eprop
\proof
We may assume $n= 2$, the general case then follows by induction. 
We have to show that
\ben\label{UV3}
\lim_{s\to 0}\frac{\Im(\Phi, Z(s)\Phi)}{s}  =  \a_1 + \a_2 \, , \quad ||Z(s)\Phi -\Phi||^2 =  o(s)\, .
\een
The following elementary identity holds
\begin{align}
(\Phi, Z(s)\Phi) -1 &=  \big((\Phi, V_1(s)\Phi) -1\big) + \big((\Phi, V_2 (s)\Phi) -1\big)  +  \big( (V^*_1(s)\Phi -\Phi, V_2(s)\Phi - \Phi)\big) \\
&=  \big((\Phi, V_1(s)\Phi) -1\big) + \big((\Phi, V_2 (s)\Phi) -1\big)  +  o(s) \, , \label{Zo}
\end{align}
where the second equality follows because
\ben\label{os2}
\big|(V^*_1(s)\Phi -\Phi, V_2(s)\Phi - \Phi)\big| \leq 
||V_1(s)\Phi -\Phi || \, ||V_2(s)\Phi -\Phi || = o(s)
\een
as $\partial_s \Re(\Phi, w_k(s)\Phi)  =0$
by  assumption.
Thus \eqref{UV3} follows by \eqref{Zo}. 
\eproof
\bcor\label{Up}
Let $A_k$, $k= 1,2,\dots n$, be selfadjoint operators on the Hilbert space $\H$, $\Phi\in\H$ a vector state such that every $(\Phi, A_k \Phi)$ is finite, and set $U_k(s) = e^{isA_k}$. 

Then
\ben\label{UV4}
-i\partial_s (\Phi, Z(s)\Phi)  = \sum_{k=1}^n (\Phi, A_k \Phi) \, ,
\een
with $Z(s) = U_1(s)U_{2}(s)\cdots U_n(s)$. 
\ecor
\proof
Combine Lemma \ref{Ad} and Proposition \ref{Up1}.
\eproof
\begin{remark}\label{rem1}
Corollary \ref{Up} still holds if $U_k(s)$ is any one-parameter family of unitaries such that $-i\partial_s (\Phi, U_k(s)\Phi) = (\Phi, A_k \Phi)$. 
\end{remark}

\subsection{Relative modular operator}
Let $\M$ be a von Neumann algebra on a Hilbert space $\H$, and $\Om,\Phi\in\H$ cyclic and separating vectors for $\M$. We consider the anti-linear {\it  relative Tomita's operators} on $\H$ given by
\[
S_{\Om,\Phi} = S^\M_{\Om,\Phi} : x\Phi \mapsto x^*\Om\, , \quad x\in \M\, ,
\]
\ben\label{SF}
S'_{\Om,\Phi} = S^{\M'}_{\Om,\Phi}: x'\Phi \mapsto {x'}^*\Om\, , \quad x'\in \M'\, .
\een
We have
\[
(x'\Phi, S_{\Om,\Phi}x\Phi) = (x'\Phi, x^*\Om) = (x\Phi, x'^*\Om) = (x\Phi, S'_{\Om,\Phi} x' \Phi)
\]
thus $S^*_{\Om,\Phi} \supset S'_{\Om,\Phi}$; indeed,
\ben\label{SF*}
S^*_{\Om,\Phi} = S'_{\Om,\Phi}\, .
\een
We use the same symbol for the operators $S_{\Om,\Phi},S'_{\Om,\Phi}$ and their closures.

By considering the polar decompositions, we get the {\it relative modular operators and conjugations}:
\[
S_{\Om,\Phi} = J_{\Om,\Phi}\Delta^{1/2}_{\Om,\Phi}\, ,\qquad S'_{\Om,\Phi} = J'_{\Om,\Phi}\Delta'^{1/2}_{\Om,\Phi}\, .
\]
From \eqref{SF*} we have
\[
J'_{\Om,\Phi}{\Delta'}^{1/2}_{\Om,\Phi} = \Delta^{1/2}_{\Om,\Phi} J^*_{\Om,\Phi} = J^*_{\Om,\Phi}(J_{\Om,\Phi}\Delta^{1/2}_{\Om,\Phi} J^*_{\Om,\Phi})\, ,
\]
thus
\ben\label{JJ'}
J'_{\Om,\Phi} = J^*_{\Om,\Phi}\, , \qquad  \Delta'_{\Om,\Phi} = J_{\Om,\Phi}\Delta_{\Om,\Phi} J^*_{\Om,\Phi} 
\een
by the uniqueness of the polar decomposition. 

On the other hand, $S_{\Phi, \Om} = S^{-1}_{\Om,\Phi}$, so
\[
S_{\Phi,\Om}  = J_{\Phi,\Om}\Delta^{1/2}_{\Phi,\Om} = S^{-1}_{\Om,\Phi}  =  \Delta^{-1/2}_{\Om,\Phi}J^*_{\Om,\Phi} 
= J^*_{\Om,\Phi} (J_{\Om,\Phi} \Delta^{-1/2}_{\Om,\Phi}J^*_{\Om,\Phi}) \, ,
\]
thus
\ben\label{JJ2}
J_{\Phi,\Om} = J^*_{\Om,\Phi} \, , \qquad  \Delta_{\Phi,\Om}  = J_{\Om,\Phi} \Delta^{-1}_{\Om,\Phi}J^*_{\Om,\Phi}\, .
\een
Putting together \eqref{JJ'} and \eqref{JJ2}, we have
\ben\label{JJ3}
J'_{\Om,\Phi} = J_{\Phi,\Om} = J^*_{\Om,\Phi}  \, , \qquad  \Delta'_{\Om,\Phi} = {\Delta}^{- 1}_{\Phi,\Om} = J_{\Om,\Phi}\Delta_{\Om,\Phi} J^*_{\Om,\Phi} \, .
\een
Now, with $\f = (\Phi, \cdot \Phi)$, $\om = (\Om, \cdot \Om)$ the states on $\M$ associated with $\Phi,\Om$, the following formulas for the Connes cocycles hold:
\ben\label{uDD}
u_s =(D\om : D\f)_s = \Delta^{is}_{\Om,\Phi}\Delta^{-is}_{\Phi}  \, , \qquad (D\f : D\om)_s = \Delta^{is}_{\Phi,\Om}\Delta^{-is}_{\Om}\, ;
\een
since $(D\om :D \f)_s = (D\f : D\om)^*_s$, we have by \eqref{JJ3}
\ben\label{uu}
u_s = \Delta^{is}_{\Om,\Phi}\Delta^{-is}_{\Phi} = \big(\Delta^{is}_{\Phi,\Om}\Delta^{-is}_{\Om}\big)^*
= \Delta^{is}_{\Om}\Delta^{-is}_{\Phi,\Om} = \Delta^{is}_{\Om}\Delta'^{is}_{\Om,\Phi}\, .
\een
Set $u'_s  =(D\om' : D\f')_s$, with $\f' = (\Phi, \cdot \Phi)$, $\om' = (\Om, \cdot \Om)$ on $\M'$. 
Then 
\[
u'_s = \Delta'^{is}_{\Om,\Phi}\Delta'^{-is}_{\Phi} = \Delta'^{is}_{\Om,\Phi}\Delta^{is}_{\Phi}\, ,
\]
so
\ben\label{uu2}
u_{-s} u'_{s} = \Delta^{-is}_{\Om}\Delta^{is}_{\Phi} \, .
\een
{For more properties of the relative modular operators and conjugations, we refer to \cite[Thm C]{AM82}.} 
\subsection{Relative entropy}
Let $\om, \f$ be faithful, normal positive linear functionals on a von Neumann algebra $\M$ on the Hilbert space $\H$. The {\it relative entropy} between $\om$ and $\f$ is defined by
\ben\label{Edef}
S(\f|\!| \om) = - (\Phi, \log \Delta_{\Om,\Phi}\Phi)\, ,
\een
with $\Om$ and $\Phi$ cyclic and separating vectors of $\H$ giving $\om$ and $\f$ and $ \Delta_{\Om,\Phi}$ is the relative modular operator as above; indeed, $(\Phi, \log \Delta_{\Om,\Phi}\Phi)$ is well-defined,
$S(\f|\!| \om)$ is non-negative and does not depend on the choice of the vector representatives $\Om$ and $\Phi$. This definition extends to the case $\om$, $\f$ are not faithful, see \cite{Ar76, Ar77}. 
 
We also recall the definition or the relative entropy in terms of the {\it spatial derivative}, see \cite{OP}. With $\om,\f' $ normal states on $\M$ and its commutant $\M'$,  Connes' spatial derivative 
$\Delta(\om/\f') = \frac{d\om}{d\f'}$ is a positive selfadjoint operator on $\H$ with support $ee'$, where $e\in \M$
is the support of $\om$ and $e'\in \M'$ is the support of $\f'$. 

Let $\f$ and $\om$  be normal (not necessarily faithful) positive linear functionals  on $\M$, and $\Phi\in\H$ any vector giving the functionals $\f$ on $\M$. Denote by $\f'$ the  functional $(\Phi, \cdot \Phi)$ on $\M'$. The relative entropy between $\om$ and $\f$ is defined by
\ben\label{Edef2}
S(\f|\!| \om)  = - (\Phi, \log \Delta(\om/\f')\Phi )\, ,
\een
if $e' < e$, with $e,e'$ as above (and $\log \Delta(\om/\f')$ defined by functional calculus on $e'\H$).    
Indeed, $\Delta(\om/\f') = \Delta_{\Om,\Phi}$, with a suitable definition of the relative modular operator in the non-faithful case. 
If $e' \nleq e$, we set $S(\f|\!| \om) = +\infty$.

The following formula holds in general:
\ben\label{Uh}
S(\f|\!| \om)  = - \partial_t \f\big(( D\om : D\f)_{-it}\big) 
\een
(right derivative), with $\f\big((D\om : D\f)_{-it}\big) = \act\f\big((D\om : D\f)_{s}\big)$ \cite[(5.3)]{OP}. 

The following formula also holds 
\ben\label{Uhf}
S(\f |\!| \om)  =  i \partial_s (\Phi, \Delta^{is}_{\Om, \Phi} \Phi)  =  i \partial_s \f\big((D\om : D\f)_s \big)  \, ,
\een
provided $ S(\f |\!| \om) < \infty$.

With $\psi_i$ normal positive functionals of $\M$ ($i= 1,2$), recall now that $\psi_1 \leq \psi_2$ means that $\psi_1(x^*x) \leq \psi_2(x^*x)$ for all $x\in \M$. If $\eta_i\in\H$ are vectors giving $\psi_i$
\[
\psi_1 \leq \psi_2  \  \Longleftrightarrow \ \exists\, m'\in \M', \ ||m'|| \leq 1\ \text{such that}\ \psi'_1 = \psi'_2(m'^*\cdot m') \, ,
\]
with $\psi_i = (\eta_i, \cdot \eta_i)|_{\M'}$;
indeed $m'$ is given by $m'x\eta_2 = x\eta_1$, $x\in\M$, and $m'$ is zero on the orthogonal complement of $\M\eta_2$. 

The following implication holds
\ben\label{of}
\psi_1 \leq \psi_2\  \Longrightarrow \ S(\f|\!| \psi_2) \leq S(\f|\!| \psi_1)
\een
with $\f$ any normal positive functional of $\M$, see
\cite[Cor. 5.12]{OP}. %
\blem\label{fo}
Let $\M$ be a von Neumann algebra with normal states $\f$, $\om$. 
If
\ben\label{foe}
 \f \leq (1 + \e) \om\, ,
\een
with  $\e\geq 0$,
then
\[
 S(\f |\!|\om) \leq \e \, .
\]
\elem
\proof
As $(1 + \e)^{-1} \f \leq  \om$ \eqref{foe}, the inequality \eqref{of} gives
\[
S(\f |\!| \om) \leq S(\f |\!| (1 + \e)^{-1} \f) = S(\f |\!| \f) + \log(1 + \e) = \log ( 1 + \e) \leq \e\, .
\]
\eproof
Given vector states $\Phi,\Om\in \H$, we set
\[
S(\Phi |\!| \Om)_\M = S(\f |\!| \om)\, ,
\] 
where $\f = (\Phi, \cdot\Phi)|_\M$, $\om = (\Om, \cdot\Om)|_\M$ are the states on $\M$ associated with $\Phi, \Om$. 
Set $\f' = (\Phi, \cdot\Phi)|_{\M'}$, $\om' = (\Om, \cdot\Om)|_{\M'}$; 
Lemma \ref{fo}, applied to $\M'$, reads 
\ben\label{leq}
(\Phi, \cdot \Phi)|_{\M'} \leq (1 +\e)(\Om, \cdot\Om)|_{\M'} \implies S(\Phi |\!| \Om)_{\M'} \leq \e\, .
\een
\bprop\label{pMM}
Let $\M$ be a von Neumann algebra and $\Phi, \Om$ cyclic and separating vectors. 

$a)$ Assume $S(\Phi |\!| \Om)_\M < \infty$, $S(\Phi |\!| \Om)_{\M'} < \infty$. Then 
\ben\label{MM0}
 S(\Phi |\!| \Om)_{\M'}   =  S(\Phi |\!| \Om)_\M - i\partial_s (\Phi, \Delta^{is}_\Om \Phi)\, ,
\een
with $\Delta_\Om$ the modular operator of $\M,\Om$. 

Thus, if in addition $(\Phi, \log\Delta_\Om \Phi)$ is finite,
\ben\label{MM}
 S(\Phi |\!| \Om)_{\M'}   =  S(\Phi |\!| \Om)_\M +(\Phi, \log\Delta_\Om \Phi)\, .
\een

$b)$ 
If $\Phi\in\P^\natural_\Om(M)$, the natural cone associated with $M,\Om$,
then 
\ben\label{MM2}
S(\Phi |\!| \Om)_{\M}  = S(\Phi |\!| \Om)_{\M'} \, .
\een
\eprop
\proof
$a)$
Set
\[
u_s = (D\om:D\f)_s\, \quad u'_s = (D\om':D\f')_s\, ,
\]
with $\f$, $\f'$ and $\om$, $\om'$ the functionals $(\Phi, \cdot \Phi)$ and $(\Om, \cdot \Om)$ on $\M$ and $\M'$. 

By the relations \eqref{uu2}, 
we have $u_{-s}u'_s = \Delta^{-is}_\Om\Delta^{is}_\Phi$, thus 
\[
u_{-s}u'_s\Phi = \Delta^{-is}_\Om \Phi \, ;
\] 
therefore,
\[
 -i \partial_s (\Phi,\Delta^{-is}_\Om\Phi)   = 
-i \partial_s (\Phi, u_{-s}u'_s\Phi)  = -S(\Phi |\!| \Om)_\M + S(\Phi |\!| \Om)_{\M'}\, ,
\]
where the last equality follows by Corollary \ref{Up}. 

Formula \eqref{MM} then follows from Lemma \ref{Ad}. 

$b)$  If $\Phi\in\P^\natural_\Om(\M)$, then $J\Phi = \Phi$  with $J \equiv J_\Om = J_{\Om,\Phi}$. 
By \eqref{JJ3}, we have 
$\Delta'_{\Om,\Phi} = J\Delta_{\Om,\Phi} J$, so
\[
S(\Phi |\!| \Om)_{\M'}  = - (\Phi, \log \Delta'_{\Om,\Phi} \Phi) = - (\Phi, J \log \Delta_{\Om,\Phi} J\Phi) 
 = - (\Phi,  \log \Delta_{\Om,\Phi} \Phi) = S(\Phi |\!| \Om)_{\M}
\]
and \eqref{MM2} holds. 
\eproof
Note that, if $\Phi$ belongs to the natural cone of $\Om$, we have 
\[
(D\om' : D\f')_s ={ \Delta'}^{is}_{\Om,\Phi}{\Delta'}^{-is}_{\Phi} = J{ \Delta}^{is}_{\Om,\Phi}{\Delta}^{-is}_{\Phi}J
= J(D\om:  D\f)_s J\, .
\]
\endproof

\subsection{Inner endomorphisms}
Let $\M$ be a von Neumann algebra on the Hilbert space $\H$. An endomorphism $\r$ of $\M$ is {\it inner} if 
there exists a family of isometries $w_k\in \M$, $k = 1, \dots n$, with $\sum_{k=1}^n w_k w_k^* = 1$ { (thus $w^*_k w_j = \delta_{jk}$)}, such that
\ben\label{Vr}
\r(x) = \sum_{k=1}^n w_k x w^*_k \, ,\quad x\in \M\, .
\een
In the following, we consider an inner endomorphism $\r$ with a finite family $\{w_k : k = 1, 2, \dots ,n\}$; the {\it dimension} of $\r$ is 
$d(\r)= n$, and is independent of the choice of the isometries implementing $\r$ (see \cite{L97} for an intrinsic definition). 

The {\it standard left inverse} of $\r$ is the completely positive map $\Psi : \M \to \M$ defined by
\[
\Psi_\r(x) = \frac1n \sum_{k = 1}^n w^*_k x w_k\, , \quad x\in\M\, ,
\]
and satisfies $\Psi_\r  \r = {\rm id}$. 

If $\om$ is a faithful normal state of $\M$, the Connes cocycle between $\om$ and $\om\Psi$ is given by
\ben\label{Cf}
(D\om\Psi_\r : D\om)_s = d(\r)^{-is} \sum_{k= 1}^n w_k\s_s^\om(w^*_k)\, , \quad s\in\RR\, ,
\een
where $\s^\om$ is the modular group of $\M$ associated with $\om$,
see \cite{I, L97}. 

If $v\in \M$ is an isometry, set $\om_v = \om(v^*\cdot v)$; then
\[
(D\om_v: D\om)_s = v \s^\om_s (v^*) \, ,
\]
\cite[Cor. 3.7]{Str},
so, by the chain rule for the Connes cocycles 
\[
(D\om\Psi_\r : D\om_v)_s = (D\om\Psi_\r : D\om)_s (D\om : D\om_v)_s
\]
we get
\[
(D\om\Psi_\r : D\om_v )_s = d(\r)^{-is} \sum_{k= 1}^n w_k\s_s^\om(w^*_k v)v^*\, ;
\]
therefore
\[
\om_v\big((D\om\Psi_\r : D\om_v)_s\big) = d(\r)^{-is} \sum_{k= 1}^n \om\big(v^*  w_k \s_s^\om(w^*_k v)\big) 
= d(\r)^{-is} \sum_{k = 1}^n(w_k^* v \Om, \Delta^{is}_\Om w_k^* v \Om)\, ,
\]
with $\Om$ the cyclic and separating vector giving $\om$ and 
$\Delta_\Om$ is the associated modular operator. 
By formula \eqref{Uh}, we then have
\begin{multline}
S(\om_v |\!| \om\Psi_\r)\label{Sv}
 =  - i\partial_t\om_v\big(( D\om\Psi_\r : D\om_v)_{-it}\big) 
 = - \partial_t \Big(d(\r)^{-t} \sum_{k= 1}^n(w_k^* v \Om, \Delta^{t}_\Om w_k^* v \Om) \Big)\\
 = -\sum_{k= 1}^n(w_k^* v \Om, (\log \Delta_\Om) w_k^* v \Om) + \log d(\r)\, ;
\end{multline}
the last equality is justified by \cite[Lemma 2.1]{L24} because $ (w_k^* v \Om, (\log \Delta_\Om) w_k^* v \Om) < +\infty$ 
by \cite[Lemma 2.3]{L24} as $w_k^* v \in \M$. 
\blem\label{foVi}
Let $\M$ be a von Neumann algebra, $\r$ an inner endomorphism of $\M$,  $d(\r) < \infty$, and $v\in \M$ an isometry as above. With $\om$ a faithful normal state of $\M$ given by the cyclic and separating vector $\Om$, we have
\[
S(\om_v\r |\!| \om) \leq  -\sum_{k= 1}^{d(\r)}(w_k^* v \Om, \log \Delta_\Om w_k^* v \Om) + \log d(\r)\, ,
\]
where $\Delta_\Om$ is the modular operator of $\Om$. 
\elem
\proof
With $\Psi_\r$ the standard left inverse of $\r$, by the monotonicity of the relative entropy, we have
\[
S(\om_v\r |\!| \om) = S(\om_v \r |\!| \om \Psi_\r\r) 
\leq S(\om_v |\!| \om\Psi_\r) 
  =  -\sum_{k}(w_k^* v \Om, (\log \Delta_\Om) w_k^* v \Om) +  \log d(\r)\, ,
\]
where the last equality is given by \eqref{Sv}. 
\eproof
In Section \ref{cal}, we shall need a more general form of Proposition \ref{pMM}. 
 
Let $\Phi\in\H$ be a cyclic and separating unit vector for $\M$ anf $\f$ the associated state of $\M$. 
By the chain rule for the Connes cocycles we have by \eqref{uDD}  and \eqref{Cf}
\begin{multline}\label{Dof}
(D\om\Psi_\r : D\f)_s = (D\om\Psi_\r : D\om)_s (D\om : D\f)_s  =\\
d(\r)^{-is} \sum_{k= 1}^n w_k\s_s^\om(w^*_k) \Delta^{is}_{\Om}\Delta^{-is}_{\Phi,\Om}
= d(\r)^{-is} \sum_{k= 1}^n w_k  \Delta^{is}_{\Om}w^*_k \Delta^{-is}_{\Phi,\Om} \, ,
\end{multline}
with $\Delta_\Om$ is the modular operator of $\Om$ and $\Delta_{\Phi,\Om}$ is the relative modular operator of $\Phi,\Om$
(equal to the spatial derivative: $\Delta_{\Phi,\Om} = \Delta(\f'/\om)$);
here $\f, \f'$ are the states on $\M,\M'$ given by $\Phi$. 

Since $\Delta^{-1}_{\Phi,\Om} = \Delta'_{\Om,\Phi}$ by \eqref{JJ3}, we thus have
\ben\label{uD}
(D\om\Psi_\r : D\f)_s \Delta'^{-is}_{\Om,\Phi} = d(\r)^{-is} \sum_{k= 1}^n w_k  \Delta^{is}_{\Om}w^*_k
\een
\bprop
With $\f = (\Phi, \cdot\Phi)|_\M$ and $\r$ be as above; assume that $S(\f |\!| \om\Phi) < \infty$, $S(\Phi |\!| \Om)_{\M'} <\infty$. Then

\ben\label{p1}
S(\f |\!| \om\Psi_\r)  =  S(\Phi |\!| \Om)_{\M'}   + i \sum_{k= 1}^n \partial_s(w_k^* \Phi, \Delta^{is}_\Om w_k^* \Phi) + \log d(\r)\, .
\een
Therefore,
\ben\label{p2}
S(\f{ \r} |\!| \om)  \leq  S(\Phi |\!| \Om)_{\M'}   + i \sum_{k= 1}^n \partial_s(w_k^* \Phi, \Delta^{is}_\Om w_k^* \Phi) + \log d(\r)\, .
\een
\eprop
\proof
From \eqref{uD}, we have
\[
(\Phi, (D\om\Psi_\r : D\f)_s \Delta'^{-is}_{\Om,\Phi}\Phi)  = d(\r)^{-is} \sum_{k= 1}^n (\Phi, w_k  \Delta^{is}_{\Om}w^*_k\Phi)\, ,
\]
therefore, by differentiating both sides of this equality taking into account the Leibniz rule in Proposition \ref{Up1}, we get
\begin{align*}
S(\f |\!| \om\Psi_\r)_{ \M}  -  S(\Phi |\!| \Om)_{\M'} &=
i\partial_s (\Phi, (D\om\Psi_\r : D\f)_s\Phi) + i\partial_s (\Phi, \Delta'^{-is}_{\Om,\Phi}\Phi) \\  
&= i\partial_s(\Phi, (D\om\Psi_\r : D\f)_s \Delta'^{-is}_{\Om,\Phi}\Phi) \\
&= i \sum_{k= 1}^n \partial_s(w_k^* \Phi, \Delta^{is}_\Om w_k^* \Phi) + \log d(\r)\, .
\end{align*}
The inequality \eqref{p2} follows from \eqref{p1} by the monotonicity of the relative entropy under completely positive maps { such as $\r$}. 
\endproof

\section{Entropy-energy bounds in QFT}\label{Appr}
We now perform our general analysis in the  QFT context. 

\subsection{Approximately localized vector states}\label{QFT1}
Let $\A$ be a net of von Neumann algebras on the Minkowski spacetime $\RR^{d+1}$, on the Hilbert space $\H$. Namely $\A$ is a isotonic map
\[
O \in \K \mapsto \A(O) \subset B(\H)
\]
from the family { $\K$} of double cones of $\RR^{d+1}$ to the von Neumann algebras on $\H$. 

If $B\subset \RR^{d+1}$ is any region, we denote by $\A(B)$ the von Neumann algebra generated by all the $\A(O)$ { with $O$ a subset of the domain of dependence of $B$} (where we set $\A(B) := \CC$ if $B$ has empty interior). 

We assume the following:

$\bullet$ {\it Translation covariance.} There exists a unitary representation $U$ on $\H$ such that $U(a)\A(O)U(a)^* = \A(O + a)$, 
for all $a\in \RR^{d+1}$, $O\in\K$. 

$\bullet$ {\it Spectrum condition.} The $U$-spectrum is contained in the closure of the forward light cone. 

$\bullet$ {\it Vacuum vector.} There exists a $U$-invariant vector $\Om$, and $\Om$ is cyclic for $\A(W)$ for every wedge region $W$.  

$\bullet$ {\it Twisted wedge duality.} There exists a $\Om$-fixing unitary $\Ga$ on $\H$, commuting with $U$, such that $\A(W)' = \Ga\A(W')\Ga^*$ for every wedge region $W$. 

As a consequence, $\Om$ is cyclic and separating for every wedge von Neumann algebra $\A(W)$ and we denote by $\Delta_W$ the vacuum modular operator of $\A(W)$. We have $\Delta_{W'} = \Ga^* \Delta^{-1}_W\Ga$. We further assume that
\ben\label{DD'}
\Delta_{W'} =  \Delta^{-1}_W\, ,
\een
that is, $\Ga$ commutes with $\Delta_{W}$. This is clearly satisfied if $\A$ is local or, more generally, if normal commutation relations holds, see \cite{GL95}. 

With $a\in \RR$, we denote by $\A_a = \A(W_a)$  the
von Neumann algebra associated with the wedge $W_a = \{x \in \RR^{d+1}: x_1 > x_0 + a\} $, and by $U(x_0, x_1) = U(x_0, x_1, 0,\dots, 0) $ the $x_0$-$x_1$
translation unitaries.  The vacuum the modular operators of $\A_a$, $\A'_a$ are denoted by 
 by $\Delta_{a}$. 

Given $a,b\in\RR$, by Borchers' theorem \cite{Bo92} we have
\ben\label{sz} 
\Delta^{is}_{b}  \Delta^{-is}_{a} 
= U(0, b)\Delta^{is}_{0} U(0, a -b) \Delta^{-is}_{0} U(0, -a)= U(\l(s)) \, ,
\een
with 
\ben\label{ls}
\l(s) = \big((b-a)\sinh 2\pi s, (a-b)\cosh 2\pi s + b- a\big)\, , \quad s\in \RR\, .
\een
We denote by
$P$ the Hamiltonian\footnote{{ Here our notation deviates for brevity from that in the Introduction and Outlook sections, where we write $P\equiv P^0$ for the generator of the translations in the $x^0$-direction.}}, i.e., the generator of the time-translation one parameter unitary group. 
\blem\label{lemP}
For every $\Phi\in\H$ with finite energy, namely $(\Phi, P\Phi) < \infty$, we have
\ben\label{aP}
-i\partial_s (\Phi, U(\l(s))\Phi)  =  2\pi (b -a) (\Phi, P\Phi)\, ,
\een
where $\l(s)$ is defined in \eqref{ls}. Moreover, if $\Phi\in D(P)$, the domain of $P$,
we have
\ben\label{aP2}
-i\partial_s U(\l(s))\Phi  =  2\pi (b -a) P\Phi
\een
(strong limit of the incremental ratio). 
\elem
\proof
We have $P = \frac12(P_+ + P_-)$,
with $P_\pm$ the null Hamiltonians in the null $x_0$-$x_1$ directions, therefore $(\Phi, P\Phi) < \infty$ implies $(\Phi, P_\pm\Phi) < \infty$. It follows that $(\Phi, U(0,s)\Phi)$ is differentiable at $s=0$, indeed $-i \partial_s (\Phi, U(0,s)\Phi) = \frac12\big((\Phi, P_+ \Phi) - (\Phi, P_-\Phi)\big)$. Then \eqref{aP} follows straightforwardly from the Leibniz rule given by Proposition \ref{Up1}. 

Formula \eqref{aP2} follows similarly as above by Stone's theorem. 
\eproof
From \eqref{sz} we have
\[
(\Phi, \Delta^{is}_{b}\Phi)  = (\Phi, U(\l(s)) \Delta^{is}_{a}  \Phi) 
\]
therefore, if $\Phi\in D(P)$, we derive from  \eqref{aP2} that
\ben\label{DDP}
-i \partial_s (\Phi, \Delta^{is}_{b}\Phi)  = 
-i \partial _s (\Phi, \Delta^{is}_{a}  \Phi) +  2\pi (b -a)  (\Phi, P\Phi)\, .
\een
We now consider the vacuum relative entropy of a vector state $\Phi$ on $\A_a$ and $\A'_a$: 
\ben\label{Sa}
S(a) = S(\Phi |\!| \Om)_{\A_{ a}}\, ,\qquad {\bar S}(a) = S(\Phi |\!| \Om)_{\A'_{ a}}\, .
\een
The following proposition is very similar to \cite[Lemma 1]{C18}, though our assumptions are different, as is the method of our proof. 
{Most significantly, here we consider the entropy difference in the spacelike direction rather than in the lightlike direction.}
\bprop\label{SSP}
Let $a,b\in\RR$ with $S(a) <\infty$, ${\bar S}(b) <\infty$. We have
\ben\label{dS1}
S(a) - S(b)   =   {\bar S}(a) - {\bar S}(b) + 2\pi(b-a) (\Phi, P\Phi) \, ,
\een
for every vector state $\Phi\in D(P)$. 
\eprop
\proof
For simplicity, we assume that $\Phi$ is cyclic and separating for the wedge von Neumann algebras, but this is not necessary. 
Clearly, we may assume $a \leq b$. 

We have $S(b) < \infty$ and $\bar S(a) < \infty$ by the monotonicity of the relative entropy. 
So  we have by Proposition \ref{pMM}, eq. \eqref{DDP} and eq. \eqref{DD'}
\begin{align*}
S(a) + \bar S(b)  +  i\partial_s (\Phi , \Delta_{b}^{is}\Phi)  &=  \bar S(a) + S(b) + 
i\partial_s (\Phi , \Delta_{a}^{is}\Phi) \\
 &=  \bar S(a) + S(b) + i \partial _s (\Phi, \Delta^{is}_{b}  \Phi) + (\Phi, P\Phi) \, ;
\end{align*}
since $i \partial _s (\Phi, \Delta^{is}_{b}  \Phi)$ is finite by our assumptions and Proposition \ref{pMM}, we then have
\[
S(a) + \bar S(b)   
=  \bar S(a) + S(b) +  2\pi (b -a) (\Phi, P\Phi) \, ,
\]
that proves our proposition. 
\eproof
Since $S(a)$ and $-{\bar S}(a)$ are decreasing functions, they are both differentiable almost everywhere. Next corollary shows that, under the hypothesis of the above proposition, the set of differentiability for $S(a)$ and ${\bar S}(a)$ is the same. 
\bcor\label{dS}
Under the hypothesis of Proposition \ref{SSP},  $\partial^\pm_a S(a)$ exists finite iff $\partial^\pm_a {\bar S}(a)  $ exists finite and in
this case
\ben
\partial^\pm_a {\bar S}(a)   - \partial^\pm_a S(a)  = 2\pi (\Phi, P\Phi)\, .
\een
\ecor
\proof Immediate by formula \eqref{dS1}, by dividing both terms by $b-a$ and letting $b \to a^\pm$. 
\eproof
\bcor\label{CorB}
Let $B\subset \RR^{d+ 1}$ be a region contained in the tube $I_R = W_{-R} \cap W'_R$, $\sr >0$, and $\Phi\in D(P)$ a vector state such that $S(-\sr) < \infty$ and  $\bar{S}(\sr) < \infty$. Then
\ben\label{Bek1}
S(\Phi |\!|\Om)_{\A(B)} \leq S(\Phi |\!|\Om)_{\A(B')} + 2\pi \sr\,(\Phi, P\Phi)\, . 
\een
\ecor
\proof
As $S(-\sr)$ and  $\bar{S}(\sr)$ are finite, also $S(\sr)$ and  $\bar{S}(-\sr)$ 
are finite by the monotonicity of the relative entropy. 
 
Since $B \subset I_R$ we have,
again by the monotonicity of the relative entropy,
\[
S(\Phi |\!|\Om)_{\A(B)}  \leq S(-\sr)\, ,\qquad S(\Phi |\!|\Om)_{\A(B)}  \leq {\bar S}(\sr)\, .
\] 
Thus, by Proposition \ref{SSP},
\begin{align}
S(\Phi |\!|\Om)_{\A(B)}  &\leq  \frac12\big(S(-\sr) + {\bar S}(\sr)\big)
\nonumber \\ 
&=  \frac12\big({\bar S}(-\sr) + S(\sr)\big) +  2\pi \sr\, (\Phi, P\Phi) \nonumber \\
&\leq S(\Phi |\!|\Om)_{\A(B')} + 2\pi \sr\,(\Phi, P\Phi)\, , \label{B4}
\end{align}
where the last inequality follows 
because $W'_{-R}, W_R \subset B'$, once more by the monotonicity of the relative entropy. So  \eqref{Bek1} holds. 
\eproof
In the following theorem, $\om' = (\Om, \cdot\Om)|_{\A(B')}$ and $\f' = (\Phi, \cdot\Phi)|_{\A(B')}$ are the vacuum state and the state given a the vector $\Phi\in\H$ on $\A(B')$. 
Of course, $\f' \leq (1 + \e)\om'$ means that $ (\Phi , x'^* x' \Phi) \leq (1 + \e)( \Om, x'^* x'\Om)$ for all $x'\in \A(B')$, see \cite{F95}. 
\bthm\label{Thme}
Let $B\subset \RR^{d+ 1}$ be a region contained in the tube $I_R$, $\sr >0$, and $\Phi\in D(P)$ a vector state such that the $\Phi$-vacuum relative entropies $S(-R) $ and ${\bar S}(R)$ in \eqref{Sa} are finite. 

If $\f'\leq (1 + \e)\om'$ on $\A(B')$, with $\e \geq 0$, then
\ben\label{Bek2}
S(\Phi |\!|\Om)_{\A(B)} \leq  2\pi \sr\,(\Phi, P\Phi) +\e\, . 
\een
In particular, if $\Phi$ localized in $B$, then
\ben\label{Bek4}
S(\Phi |\!|\Om)_{\A(B)} \leq 2\pi \sr\,(\Phi, P\Phi)\, . 
\een
\ethm
\proof
By Corollary \ref{CorB}, we have
\[
S(\Phi |\!|\Om)_{\A(B)} 
\leq S(\Phi |\!| \Om)_{\A(B')} + 2 \pi \sr\, (\Phi, P\Phi) \leq  \e + 2\pi \sr\, (\Phi, P\Phi)
\]
as $S(\Phi |\!| \Om)_{\A(B')} \leq \e$
by Lemma 
\ref{fo} (with $\M = \A(B')$). 

The bound  \eqref{Bek4} follows from  \eqref{Bek1} since $S(\Phi |\!|\Om)_{\A(B')} = 0$ if the vector state $\Phi$ is localized in $B$. 
\eproof

\begin{remark} 
\label{rem3.6}
It was conjectured in \cite[Sect. II.4]{CT77} that, if $\M$ is factor of type $III_1$ in Connes' classification, given $\e >0$ and faithful normal states $\f, \om$ on $\M$, there exists a unitary $u\in \M$ such that 
\ben\label{CT}
(1 - \e)\om \leq \f_u \leq (1 + \e)\om 
\een
on $\M$, where $\f_u = \f(u^*\cdot u)$  (a weak version of \eqref{CT} with the norm distance was then proved in \cite{CS78}). The same would then hold for a general $III_1$ von Neumann algebra, i.e. direct integral of type $III_1$ -factors by a fiberwise application of \eqref{CT}. 

In an asymptotically scale invariant theory, the von Neumann algebra $\A(B)$, thus $\A(B')$, is of type $III_1$ \cite{F85}. If
we assume that \eqref{CT} holds, Theorem  \ref{Thme} then implies the following:
\end{remark}

\noindent
\emph{Let $B$ be a region contained in the tube $I_R$, $\sr >0$, and $\f = (\Phi, \cdot\Phi)$ a state on $\A(B)$
such that the $\Phi$-vacuum relative entropies $S(-R) $ and ${\bar S}(R)$ are finite.
Then, given $\e >0$, there exists a vector $\eta\in \H$ with $\f = (\eta, \cdot\eta)$ on $\A(B)$ such that 
\[
S(\f |\!|\om)_{\A(B)} \leq 2 \pi \sr\,(\eta, P\eta) + \e \, .
\]
}

\subsection{Charged localized states}\label{QFT2}
We aim to consider now localized states carrying a short range charge, see \cite{DHR1};
a version of the Bekenstein bound \cite{Ca08} with the boost Hamiltonian in the presence of such a charge has been studied in \cite{L18,L20}.
In this section, $\A$ is a net of von Neumann algebras with the properties as in Section \ref{QFT1}, but we assume $\A$ to be local, that is $\Ga = 1$.  

We further assume Haag duality for Minkowski convex regions $B$ (regions that are intersection of wedge regions):
\ben\label{HD}
\A(B) = \A(B')';
\een
however, this property is non necessary. Indeed we may pass to the dual net $\A^{\rm d}$
\[
\A^{\rm d}(B) = \A(B')'\,  ,
\]
and $\A^{\rm d}$ satisfies \eqref{HD} by the assumed wedge duality property. The Minkowski convex regions that will be considered are double cones and tubes. 

If $B\subset \RR^{1 + d}$ is any region, we denote by $\mathfrak A(B)$ the C$^*$-algebra generated by the von Neumann algebras $\A(O)$ as $O\in\K$ $O\in \K$ runs over the double cones $O\subset B$.  We set $\A(B) = \mathfrak A(B)''$ and 
denote by $\mathfrak A = \mathfrak  A(\RR^{d+1})$ the {\it quasi-local $C^*$-algebra}. 

A DHR {\it representation} of $\mathfrak A$ is a  representation $\r$  of $\mathfrak A$ on a Hilbert space $\H_\r$ such that 
\[
\r |_{\mathfrak A(O')} \approx {\rm id} |_{\mathfrak A(O')}\, \quad \text{for every}\ O\in\K\, ;
\]
here, ${\rm id}$ denotes the identity (vacuum) representation of $\mathfrak A$, and the symbol $\approx$ stays for unitary equivalence. We also use the symbol $\preceq$ for unitary equivalence to a subrepresentation. 

We denote by $\L_\r$ the set of states associated with the DHR representation $\r$; thus, $\f\in \L_\r$ if there is a vector $\Phi\in\H_\r$ such that
\ben\label{fp}
\f(x) = ( \Phi, \r(x)\Phi) \, .
\een
Given $\f \in \L_\r$, we may restrict $\r$ to the 
the invariant subspace $\ov{\pi(\mathfrak A)\Phi}$ of $\H_\r$ and get a DHR subrepresentation $\r_1$ of $\r$ such that $\Phi$ is cyclic for 
$\r_1$. The representation \eqref{fp} still holds with $\r_1$ in place of $\r$. We may thus assume that $\Phi$ is cyclic for $\r$; the triple 
$(\H_\r, \r, \Phi)$ representing $\f$ as in \eqref{fp} is then unique up to unitary equivalence by the uniqueness of the GNS representation. 

We shall say that the {\it state $\f\in\L_\r$ is {\it localized} in the region $B\subset \RR^{1+d}$} if $\f|_{\mathfrak A(B')} = \om|_{\mathfrak A(B')}$ with $\om$ the vacuum state of $\mathfrak A$. 

An {\it endomorphism} $\r$ of $\mathfrak A$ is {\it localized} in $B$, 
such that $\H_\r = \H$ and 
if $\r|_{\mathfrak A(B')} = {\rm id}|_{\mathfrak A(B')}$. Note that an endomorphism of $\mathfrak A$ is a representation of $\mathfrak A$ on $\H$. 
Given $O\in\K$, a DHR representation is unitary equivalent to an endomorphism of $\mathfrak A$ localized in $O$ (by Haag duality); in other words, a DHR representation is localisable in any double cone.\footnote{One may consider a representation localisable in some double cone. In this case, our bound holds if both the state $\f\in\L_\r$ and the charge $\r$ are localized in $B$.}

In the following, $B$ is a Minkowski convex region. Lemma \ref{lmin} is a local version of \cite[Thm. 2.10]{L18}. 
\blem\label{lmin}
Let $\r$ be a DHR representation of $\mathfrak A$. A state
$\f\in\L_\r$ is localized in $B$ if and only if there exists an endomorphism $\r_1$ of $\mathfrak A$ localized in $B$, $\r_1 \preceq \r$, and an isometry $v\in \A(B)$ such that 
\ben\label{fV}
\f = \om\big(v^*\r_1(\cdot)v\big)\, .
\een
We may choose $\r_1$ and $v$ such that the left support of $\r_1(\mathfrak A)v$ is equal to $1$, i.e., $\ov{\r_1(\mathfrak A)v\H} = \H$ (\emph{minimal choice}). If $\r_1', v'$ is a second minimal choice, then there exists a unitary $u\in\A(B)$ such that $v' = uv$, $\r_1' = u\r_1(\cdot)u^*$. 

If $\f = (\Phi, \r(\cdot)\Phi)$ with $\Phi$ cyclic for $\r$, then $\r_1 \approx \r$. 
\elem
\proof
Clearly, a state of $\f$ of the form \eqref{fV} belongs to $\L_\r$ and is localized in $B$. 

Conversely, assume that $\f\in\L_\r$. 
Choose an endomorphism $\r_2$ of $\mathfrak A$ localized in $B$, $\r_1 \approx \r$.  
By unitary equivalence, there exists $\Phi\in \H$ such that $\f(x) = (\Phi, \r_2(x)\Phi)$, $ x\in \mathfrak A$. 
By considering the $\Phi$-cyclic subrepresentation $\r_1 \preceq \r_2$ as above, we get an endomorphism $\r_1$ of $\mathfrak A$ localized in $B$,  such that $\f  = (\Phi, \r_1(\cdot)\Phi)$. 

As $\f$ is localized in $B$, the isometry $v$ on $\H$ given by $vx\Om = x\Phi$, $x\in \mathfrak A(B')$, belongs to $\A(B')'$, thus $v\in\A(B)$ if Haag duality holds for $B$. 

Clearly, $\f(x) = (v\Om, \r_1(x) v\Om)$, $x\in \mathfrak A$, that is eq. \eqref{fV} holds, and 
$\ov{\r(\mathfrak A)v \H} \supset \ov{\r(\mathfrak A)\Phi} = \H$ by the cyclicity of $\Phi$, that is, $(\r_1,v)$ is a minimal choice. 

We now show that the minimal choice of \eqref{fV} is unique as in the statement. Indeed, if $v_1', \r_1'$ is a second minimal choice, 
then $\r_1$ and $\r_1'$ are unitarily equivalent by the above discussion; indeed, by the uniqueness of the GNS representation, there exists a unitary $u$ such that $\r_1' = u \r_1(\cdot) u^*$, thus $u$ belongs to  $\A(B)$ by Haag duality and $uv\Om = v'\Om$, thus $uv = v'$ because $uvx\Om = x uv\Om =  xv'\Om = v'x\Om$, $x\in \A(B')$ and $\Om$ is cyclic for $\A(B')$. 

Finally, in the above, $\r_1 = \r_2 \approx \r$ if $\Phi$ is cyclic for $r$. 
\eproof
Let $\r$ be a DHR representation of $\mathfrak A$. Recall that that the {\it statistical dimension} $d(\r)$ of $\r$ is defined in \cite{DHR1}. In spacetime dimension greater than two, $d(\r)$ is either a positive integer or $+\infty$. $d(\r)$ is the square root of the Jones index of any localized endomorphism equivalent to $\r$ \cite{L89}. $\log d(\r)$ has a free energy or entropy meaning \cite{L97,L01}. 

If $\f\in\L_\r$, we may define the {\it dimension} $d(\f)$ of $\f$ by setting 
\[
d(\f) = d(\r_1)\, ,
\]
where $\r_1$ is in the minimal choice in the lemma \ref{lmin}.  

In the following, all DHR representations $\r$ are assumed to be {\it translation covariant}. 
That is, there exists a unitary, positive energy, unitary representation $U_\r$ on $\RR^{1+d}$ such that
\[
\r\big(U(a)xU(a)^*\big) = U_\r(a)\r(x)U_\r(a)^*\, ,\quad a\in \RR^{1+d}\, , \ x\in\mathfrak A\, .
\]
(See \cite{GL92} for general conditions that ensure translation covariance.)
The generator $P_\r$ of the time-translation unitary one-parameter subgroup of $U_\r$ if the (charged) {\it Hamiltonian} in the representation $\r$. 
\blem\label{Thminn}
Let $\r$ be an inner endomorphism of $\mathfrak A$ a region localized in the region $B\subset \RR^{1+d}$ of width $R$, and $\f\in\L_\r$ a state of $\mathfrak A$ localized in $B$.  We have
\[
S(\f |\!| \om)_{\A(B)}  \leq 2\pi \textit{\small R}\, (\Phi, P_\r \Phi) + \log d(\r)\, ,
\]
with $\Phi\in\H_\r$ the vector state giving $\f$ \eqref{fp}. 
More generally, this holds if $\A$ is twisted local rather than local. 
\elem
\proof
By assumption, there exist an isometry $v\in \A(B)$ and isometries $w_k\in\A(B)$ implement $\r$ as in \eqref{Vr} such that 
\[
\f = \om_v \r = (v\Om, \r(\cdot)v\Om) = \sum_k (w_k^*v\Om, \cdot w_k^*v\Om) \, .
\]
By Lemma \ref{foVi}, we have
\[
S(\f |\!| \om) \leq  -\sum_{k = 1}^n(w_k^* v \Om, (\log \Delta_{B})\, w_k^* v \Om) + \log d(\r)\, ,
\]
where $\Delta_{B}$ is the vacuum modular operator of $\A(B)$. 

By Corollary 2.11 of \cite{L24}, we have
\[
\log \Delta_{B} \leq 2 \pi \textit{\small R}\, P\, ,
\]
with $P$ the vacuum Hamiltonian of $\A$. Therefore,
\[
S(\f |\!| \om) \leq   2 \pi \textit{\small R}\sum_{i= k}^n(w_k^* v \Om,P w_k^* v \Om) + \log d(\r)
= 2 \pi \textit{\small R}\,(\Phi ,P_\r \Phi) + \log d(\r)\, ,
\]
with $P_\r = \sum_k w_k P w_k^*$ the Hamiltonian of $\A$ in the representation $\r$ and $\Phi = v\Om$. 
\eproof
In the next theorem, we assume that the space dimension $d$ is greater than one. We are going to use the Doplicher-Roberts reconstruction theorem \cite{DR}, { which states that t}here exists a {\it field net} of von Neumann algebras $\F(O)$ associated with the tensor category of DHR endomorphisms of $\mathfrak A$. Namely, there exists a Hilbert space $\tilde \H \supset \H$, a net of von Neumann algebra 
\[
O\in \K\mapsto \F(O)\subset B(\tilde\H)\, ,
\]
and a compact group $G$ of internal symmetries of $\F$ such that $\A(O)$ is (identified with) the fixed-point algebra of $\F(O)$ under the action of $G$. The net $\F$ is twisted local, with normal commutation relations. $\F$ satisfies twisted wedge duality (this follows because, if $W$ is a wedge, the action of $G$ on $\F(W)$ has full spectrum, cf. \cite{DR}). 

The identity (vacuum) representation ${\rm id}_{\mathfrak F}$ of $\mathfrak F$ restricts to $\mathfrak A$ as
\ben\label{FA}
{\rm id}_\F |_{\mathfrak A} \approx \bigoplus_\r d(\r)\r \, ,
\een
direct sum over all DHR irreducible representations $\r$ of $\A$ with finite statistics, with multiplicity. We may suppose that all the $\r$'s are endomorphisms of $\mathfrak A$ localized in a given region $B$. 
The unitary, covariance spacetime translation representation $U_\F$ of $\F$ on $\tilde \H$ decomposes as
\ben\label{FAU}
U_\F = \bigoplus_\r d(\r)U_\r
\een
accordingly to \eqref{FA}. We denote by $P_\F = \bigoplus_\r d(\r)P_\r$ the Hamiltonian of $\F$. 
\bthm\label{Bch2}
Let $\r$ be a DHR representation of $\mathfrak A$ and $\f\in\L_\r$ be a state of $\mathfrak A$ localized in a region $B\subset \RR^{1+d}$ of width $R$. Then
\[
S(\f |\!| \om)_{\A(B)}  \leq 2\pi \textit{\small R}\, (\Phi, P_\r \Phi) + \log d(\f)\, ,
\]
with $\Phi\in\H_\r$ the vector state giving $\f$ \eqref{fp}. 
\ethm
\proof
Clearly, we may assume that $d(\r)$ is finite, $\r$ is an endomorphism of $\mathfrak A$ localized in $B$, and $\Phi$ is cyclic for $\r$; thus 
$\r$ is a finite direct sum of endomorphisms appearing in \eqref{FA}. 

For simplicity, we further assume that $\r$ is irreducible, the reducible case follows along the same lines. We may then identify $\Phi$ with a vector $\tilde\Phi\in\tilde\H$ 
supported in $\H_\r\oplus \H_\r\oplus \cdots\oplus\H_\r$ ($d(\r)$-times), embedded in $\tilde \H$ by \eqref{FA}. That is, we consider the vector
\ben\label{tx}
\tilde\Phi = d{(\r)}^{-1/2} (\Phi\oplus\Phi\oplus \cdots \oplus\Phi) 
\een
of $\H_\r\oplus \H_\r\oplus \cdots\oplus\H_\r$.
Then
\[
\f(x) = (\Phi, \r(x)\Phi) = (\tilde\Phi, (\r\oplus\r\oplus\cdots\oplus\r) (x)\tilde\Phi) =
(\tilde\Phi, {\rm id}_{\mathfrak F}(x)\tilde\Phi)\, ,\quad x\in\mathfrak A \, ;
\]
so, the state $\tilde\f$ on $\mathfrak F$ given by $\tilde\f(x) = (\tilde\Phi, {\rm id}_{\mathfrak F}(x)\tilde\Phi), \, x\in \mathfrak F$, extends $\f$; and $\tilde\f$ is localized in $B$. 

Now, $\r$ extends to an inner endomorphism $\tilde \r$ of $\mathfrak F$ localized in $B$. Let $V_1, V_2,\dots V_{d(\r)}$ be isometries in $\F(B)$ implementing $\tilde \r$, so
\[
\tilde\r (x) = \sum_{k =1}^{d(\r)} V_k x V_k^*\, .
\] 
$\tilde\r$ commutes with $\epsilon$,  with $\epsilon: \mathfrak F \to \mathfrak A$ the vacuum preserving conditional expectation. 
Then $\tilde\om = \om \epsilon$ and $\tilde\f = \f \epsilon$, with
$\tilde \om$ the vacuum state of $\mathfrak F$, so $\tilde \f =\tilde\om\tilde\r$.  
We now apply the Bekenstein-type inequality to $\tilde \f$;  we have
\[
S(\f |\!| \om)_{\A(B)} = S(\tilde\f |\!| \tilde\om)_{\F(B)} 
\leq 2\pi \textit{\small R}\, (\tilde \Phi, P_\F \tilde \Phi) + \log d(\r) =  2\pi \textit{\small R}\,  ( \Phi, P_\r \Phi) + \log d(\r)  \, ,
\]
where $P_{\F}$ is the vacuum Hamiltonian of $\F$, thus $P_{\F}|_{\H_\r} = P_\r$;
the first equality holds because both states $\tilde \f$ and $\tilde \om$ preserve $\epsilon$, and the second inequality is due to Lemma \ref{Thminn}. 
\eproof

\begin{remark}
{If $\r = \id$ is the vacuum representation, 
Theorem \ref{Bch2} reduces to \cite[Thm 3.2]{L24}. On the other hand, 
the latter theorem cannot be fully derived by
Theorem \ref{Thme} (case $\e = 0$) because of the finite entropy assumptions.}
\end{remark} 

\subsection{Charged, approximately localized states}\label{cal}
We sketch here how to merge the analysis in Sections \ref{QFT1}, \ref{QFT2}, by combining the arguments therein, and consider approximately localized states that carry a localized charge. 

Let $\A$ be a local net of von Neumann algebras as in Section \ref{QFT1} ($\Ga = 1$), $\r$ a DHR representation of $\mathfrak A$ localized in a region $B\subset \RR^{1+d}$ of width $R>0$, and $\f\in\L_\r$ a state; thus $\f = (\Phi, \r(\cdot)\Phi)$, with $\Phi\in \H_\r$. 

Note that, since $\r$ is localisable in any double cone, $\r$ is normal on $\mathfrak A(E)$ if $E$ is in a region such that $E'$ has non-empty interior, thus $\f$ is also normal on $\mathfrak A(E)$; we may then consider the relative entropy $S(\f |\!| \om)_{\A(E)}$ between the restrictions of $\f$  and the vacuum state $\om$ $\mathfrak A(E)$. Set $S(a) = S(\f |\!| \om)_{\A_a}$, $\bar S(a) = S(\f |\!| \om)_{\A'_a}$, $ a\in \RR$. 
\bthm\label{Thme2}
Let the region $B\subset \RR^{d+ 1}$ be contained in the tube $I_R$, $\sr >0$, 
and $\f\in \L_\r$ as above with finite relative entropies $S(-R)$ and ${\bar S}(R)$. 

If $\f\leq (1 + \e)\om$ on $\mathfrak A(B')$, with $\e \geq 0$, then
\ben\label{Bek3}
S(\f |\!|\om)_{\A(B)} \leq  2\pi \sr\,(\Phi, P_\r\Phi) + \log d(\f) +\e\, . 
\een
\ethm
\proof
The proof combines the proofs of Theorem \ref{Thme} and Theorem \ref{Bch2}. With the notations in the proof of Theorem \ref{Bch2}, note that $\tilde\f\leq (1 + \e)\tilde\om$ on $\mathfrak F(B')$; indeed, $\epsilon$ leaves invariant both $\tilde\f$ and $\tilde\om$, so
\[
\tilde \f(x^* x)  = \f(\epsilon(x^*x)) \leq (1 + \e)\om(\epsilon(x^*x)) = (1 + \e)\tilde\om(x^*x)\, ,\quad x \in \mathfrak F(B')\, .
\]
Then by \eqref{p2} and reasoning as in the previous sections, we have
\begin{align*}
S(\f |\!| \om)_{\A(B)} &= S(\tilde\f |\!| \tilde\om)_{\F(B)}\\
& \leq  S(\tilde\f |\!| \tilde\om)_{\F(B)'} + 2\pi \textit{\small R}\,  \sum_{k= 1}^n \partial_s(V_k^*\tilde  \Phi, P_\F V_k^*\tilde  \Phi) + \log d(\r)
 \\
& \leq  \e + 2\pi \textit{\small R}\,  \sum_{k= 1}^n (V_k^*\tilde  \Phi, P_\F V_k^*\tilde  \Phi) + \log d(\r) \\
 &=  \e + 2\pi \textit{\small R}\,  ( \Phi, P_\r \Phi) + \log d(\r)  \,   .
\end{align*}
\eproof

\section{Outlook}
\label{outlook}
We end this paper by indicating some potentially fruitful extensions of our analysis. 

1) A CFT is of course a special kind of QFT with conformal invariance, so obviously, all results in this paper apply to CFTs as well. 
However, due to the extra symmetries of a CFT, we expect that stronger
energy-entropy bounds should exist. We now outline how one might obtain these in a heuristic manner, leaving aside domain issues for the various unbounded operators. 
By the Hislop-Longo theorem (see e.g., \cite{H}),  the modular operator of the vacuum state associated with an open ball $B_R$ of radius $R$ is given in a CFT by
\begin{equation}
    \log \Delta_{\Omega, R} = 
    \pi \left( RP^0 + \frac{1}{R}K^0 \right),
\end{equation}
where $K^0$ is the generator of special conformal transformations in the time-direction orthogonal to the time-slice hosting $B_R$. Let $A_{R_+,R_-} := B_{R_+} \setminus \overline{B_{R_-}}$
be an ``annulus''\footnote{$A_{R_+,R_-}$ is properly speaking an annulus only in two spatial dimensions. In one spatial dimensions, it is the disjoint union of two intervals, whereas in $d>2$ spatial dimensions, it is a spherical shell of width $R_+-R_-$.} in our time-slice, enclosed between spherical shells of radii $R_+>R_->0$, and let $B$ be any open domain contained in $A_{R_+,R_-}$. For a localized vector $\Phi=v\Omega$ in $B$ [i.e., $v$ is an isometry from $\A(B)$], we have the chain of inequalities
\begin{equation}
\begin{split}
    (R_++R_-) S(\Phi |\!| \Omega)_{\A(B)} &\le (R_++R_-) S(\Phi |\!| \Omega)_{\A(A_{R_+,R_-})} \\
    & \le 
    R_+ S(\Phi |\!| \Omega)_{\A(B_{R_+})} + 
    R_- S(\Phi |\!| \Omega)_{\A(B_{R_-})'}\\
    & = R_+ (\Phi, (\log \Delta_{\Omega, R_+}) \Phi) +
    R_- (\Phi,  (\log \Delta_{\Omega, R_-}') \Phi)\\
    & = \pi (\Phi, (R_+^2 P^0 + K^0) \Phi) - \pi (\Phi, (R_-^2 P^0 + K^0) \Phi) \\
    & = \pi (R_+^2 - R_-^2) (\Phi, P^0 \Phi).
\end{split}
\end{equation}
In the first and second line, we used the monotonicity of the relative entropy, in the third line we used Proposition \ref{pMM}, and in the fourth line we used the Hislop-Longo theorem, as well as $\Delta_{\Omega,R}'=\Delta_{\Omega,R}^{-1}$. Dividing by $R_++R_-$ we therefore have 
\begin{equation}
\label{bekenstein4}
S(\Phi |\!| \Omega)_{\A(B)} \le 
\pi(R_+-R_-) (\Phi, P^0 \Phi)
\end{equation}
for any vector state $\Phi$ localized in 
an open region $B$ contained in the annulus $A_{R_+,R_-}$. One can presumably also obtain bounds generalizing \eqref{bekenstein4} for charged, approximately localized states analogous to Theorem \ref{Thme2}.

Note that we recover the previous Bekenstein bound \eqref{bekenstein1} from \eqref{bekenstein4} in the limit as $R_\pm \to \infty$ keeping $R_+-R_-=2R$ fixed because the annulus $A_{R_+,R_-}$ converges to the strip $I_R$ of width $2R$. Thus, the CFT bound \eqref{bekenstein4}  is about as sharp as the general QFT bound \eqref{bekenstein1} for spindle-like, elongated regions $B$ fitting into a tubular region, but the CFT bound is generally sharper for thin shell-like regions $B$ fitting into an annulus. Similar remarks would presumably apply to charged, approximately localized states as in Theorem \ref{Thme2}.

In holography, there is a well-known relationship between relative entropies in the boundary CFT and a bulk relative entropy associated with some sort of entangling minimal surface anchored on $\partial B$ \cite{maldacena}. It might be interesting to see if our CFT Bekenstein bound \eqref{bekenstein4} has a holographic interpretation in this setting, e.g., if it corresponds to some sort of provable area-energy inequality in the classical dual gravity theory at large $N$. 

\smallskip

2) Another natural project, presumably of mainly technical nature, would be to generalize the results of Theorems \ref{Bch2} and \ref{Thme2} to $d=1$
spatial dimensions. Here, the DR reconstruction theorem \cite{DR}, used in our proofs, no longer applies and one may have sectors $\rho$ with braid group statistics and non-integer statistical dimensions $d(\rho)$ \cite{L89}.  See also \cite{BLPS} for an interpretation of the Bekenstein bound related to  uncertainty relations, and see \cite{Hollands:2020owv} for the interplay between entropic uncertainty relations and the Jones index (square of the statistical dimension \cite{L89}). Furthermore, free or superrenormalizable QFTs in $d=1$ spatial dimensions such as the $\phi^4_2$-models are well-known to have a mathematically well-defined formulation in the ``Schr\" odinger picture''. In this formulation, so-called ``log-Sobolev-'' inequalities have been established, see e.g., \cite{Cipriani} and references therein. 
Like the inequalities derived in this paper, the log-Sobolev inequalities relate a relative entropy to an energy. It would be interesting to see if there are deeper connections beyond this basic observation.  

\smallskip

3) A more daring, though rather vague, project would be to combine the Bekenstein bound \eqref{bekenstein1} or its relarives studied in this paper with Thorne's ``hoop conjecture'' \cite{Thorne}, a loosely formulated proposal of a necessary and sufficient condition for the formation of a horizon in non-spherical gravitational collapse. The hoop conjecture involves the notion $C$ of a circumferential radius of a spatial region $B$ and states that a horizon will be present within $B$ if the mass within $B$ exceeds that of a Schwarzschild black hole with horizon radius $C$. 

Although the notion of circumferential radius $C$ of a region $B$ may be defined in different ways (see, e.g., \cite{Flanagan} for a critical review), it is supposed to express a kind of maximum over the radii of $B$ in all possible directions. On the contrary, $R$ in \eqref{bekenstein1} is by definition more like a minimum radius, so, whatever the precise definition of both quantities in a general time-slice in a general spacetime, we might expect that $R \lesssim C$ and $4\pi C^2 \lesssim A$ (the area of $\partial B$) for reasonable definitions. Likewise, though the mass of a gravitational system within a region $B$ may be defined in different ways, we loosely identify it with $(\Phi, P^0 \Phi)$ in a semiclassical setting. Given that the horizon radius of a Schwarzschild black hole is $2G_NM$ in four dimensions\footnote{The hoop conjecture is generally expected {\it not} to hold in dimensions $>4$, because non-spherical black objects such as rings may be expected to form in this case from solenoid-like mass concentrations which are elongated in one or more directions rather than compact in all directions.}, a necessary condition for the absence of a horizon in $B$ in a semiclassical state $\Phi$ is expected in view of the Bekenstein bound to be
\begin{equation}
\label{bekenstein2}
S_B(\Phi |\!| \Omega) \lesssim \frac{A}{4G_N}, 
\end{equation}
where the surface area $A$ refers to the geometry corresponding to $\Phi$ via the semiclassical Einstein equation, $G_{\mu\nu} = 8\pi G_N (\Phi, T_{\mu\nu} \Phi)$.

A proof or even semi-rigorous discussion of \eqref{bekenstein2} would require a proper definition of a ``vacuum-like'' state, $\Omega$, in a general curved spacetime (one possibility might be a ``local vacuum state'' defined with respect to a past directed non-expanding lightsheet emanating from $\partial B$ in the spirit of \cite[Sec. 6]{hollandsthesis}), a precise  sense in which $\Phi,\Omega$ are supposed to satisfy the semiclassical Einstein equation, and a precise notion of mass inside a region in a semiclassical geometry\footnote{In a general curved spacetime geometry, the relationship between $T_{\mu\nu}$ or its quantum expectation value, and the mass of the gravitational field inside a region is not at all obvious.}, all of which are challenging problems for future work. 

We note, however, that the ``consequence'' \eqref{bekenstein2} of the hoop conjecture and the Bekenstein bound \eqref{bekenstein1} is similar in spirit to Bousso's entropy bound \cite{Bousso}, which may be seen as a covariant generalization of the Bekenstein bound. As \eqref{bekenstein1}, the bound \eqref{bekenstein2} involves the relative entropy, not the entropy as in \cite{Bousso}.

\medskip

\noindent
{\bf Acknowledgements.} 
R.L. acknowledges the MIUR Excellence Department Project awarded to the Department of Mathematics, University of Rome Tor Vergata, CUP E83C18000100006.

\noindent
{\bf Data availability.} Data sharing not applicable to this article as no data sets were generated or analysed during the current study.

\medskip

\noindent
{\bf Declarations}

\medskip

\noindent
{\bf Conflict of interest.} The author has no relevant financial or non-financial interests to disclose. 

\end{document}